\documentclass[APS,PRB,twocolumn,showpacs]{revtex4}
\usepackage{amssymb}
\usepackage{txfonts}
\usepackage{graphicx}

\begin{document}

\title[]{Transport properties, upper critical field and anisotropy of Ba(Fe$_{0.75}$Ru$_{0.25}$)$_2$As$_2$ single crystals}

\author{Jie Xing$^1$, Bing Shen$^2$, Bin Zeng$^2$, Jianzhong Liu$^1$, Xiaxin Ding$^1$,  Zhihe Wang$^1$, Huan Yang$^1$ and Hai-Hu Wen$^1$}\email{hhwen@nju.edu.cn}

\affiliation{$^1$ Center for Superconducting Physics and Materials,
National Laboratory of Solid State Microstructures and Department of
Physics, Nanjing University, Nanjing 210093, China}

\address{$^2$ National Laboratory for Superconductivity, Institute of Physics, Chinese Academy of
Sciences, P.O.Box 603, Beijing 100190, China}

\begin{abstract}

The temperature and angle dependent resistivity of
Ba(Fe$_{0.75}$Ru$_{0.25}$)$_2$As$_2$ single crystals were measured
in magnetic fields up to 14 T. The temperature dependent
resistivity with the magnetic field aligned parallel to c-axis and
ab-planes allow us to derive the slope of d$H_{c2}^{ab}/dT$ and
d$H_{c2}^c/dT$ near $T_c$ yielding an anisotropy ratio
$\Gamma=dH_{c2}^{ab}/dT/dH_{c2}^c/dT \approx$  2. By scaling the
curves of resistivity vs. angle measured at a fixed temperature
but different magnetic fields  within the framework of the
anisotropic Ginzburg-Landau theory, we obtained the anisotropy in
an alternative way. Again we found that the anisotropy
$(m_c/m_{ab})^{1/2}$ was close to 2. This value is very similar to
that in Ba$_{0.6}$K$_{0.4}$Fe$_2$As$_2$ (K-doped Ba122) and
Ba(Fe$_{0.92}$Co$_{0.08}$)$_2$As$_2$ (Co-doped Ba122). This
suggests that the 3D warping effect of the Fermi surface in
Ru-doped samples may not be stronger than that in the K-doped or
Co-doped Ba122 samples, therefore the possible nodes appearing in
Ru-doped samples cannot be ascribed to the 3D warping effect of
the Fermi surface.

\end{abstract}

\pacs{74.25.Fy, 74.25.Op, 74.70.-b}
\maketitle

The discovery of high temperature superconductivity in the iron
pnictides and chalcogenides greatly stimulates the interests in
understanding the novel pairing mechanism for superconductivity.
Although many experiments have been done and the results
demonstrate that the iron-based superconductors belong to a family
with unconventional pairing
mechanism\cite{Mazin,Kuroki,Hanaguri,Christianson,Keimer},
however, the detailed pairing mechanism and gap structure remain
unclear. The inelastic neutron scattering indicates a resonance
peak at the momentum ($\pi$, $\pi$), which strongly suggests that
the antiferromagnetic (AF) spin fluctuation may be the media for
the pairing. This picture gets partial support from the NMR
measurements in which a strong diverging of the 1/$T_1T$ is
observed when approaching the Neel
temperature\cite{ImaiNMR,IshidaNMR}. The superconductivity
vanishes simultaneously with the missing of the divergence of
1/$T_1T$ in the overdoped region, suggestive of a close
relationship between superconductivity and the AF spin
fluctuations. According to the tight-bind fitting to the band
structures, the pairing interaction by the AF spin fluctuation has
been calculated for the five orbitals. It was predicted that,
although the pairing channel is mainly the S$\pm$, while some
accidental nodes may exist on part of the Fermi
surfaces\cite{GraserPRB2009,KurokiPRB,HirschfeldPRB2010,HirschfeldReview}.
In this regard, many interesting results have been obtained. For
example, in LaFePO and KFe$_2$As$_2$, nodal gaps are inferred from
the penetration depth measurements\cite{Hicks,Carritong} and
thermal conductivity measurements\cite{LiSYKFe2As2}. Meanwhile in
many other systems, full gaps, sometime with a strong gap
anisotropy were
discovered\cite{ZengB,MatsudaPDoping,FengDLPDoping}. Recently,
another interesting system, namely,
Ba(Fe$_{1-x}$Ru$_{x}$)$_2$As$_2$ has received intensive
investigations\cite{Sharma,Thaler,Goldman,Poissonnet}. Studies  by
thermal conductivity on Ba(Fe$_{1-x}$Ru$_{x}$)$_2$As$_2$
\cite{LiSY} revealed that nodal superconducting gap may exist in
this material. Angle-resolved photo-emission spectroscopy (ARPES)
indicates that the band structure seems not changing too much when
crossing the wide underdoped region, leading to the great concern
about the nesting effect as the pairing driven
force\cite{Kaminski}. For a gap with the structure of cos $k_x
\cdot$ cos $k_y$ or cos $k_x$+cos $k_y$ (the two main gap
functions in the iron-based superconductors), the strong warping
effect of the Fermi surfaces may lead to an intersect of the
zero-gap line and the 3D fermi surface, resulting in horizontal
nodal lines\cite{FengDLPDoping,Prozorov,Reid}. If the possible
nodal gaps are induced by the strong warping effect of the 3D
Fermi surface, the anisotropy measured by
$\Gamma$=$H^{ab}_{c2}/H^{c}_{c2}$ should be very different in the
present system compared with other optimally doped 122 systems,
such as Ba$_{0.6}$K$_{0.4}$Fe$_2$As$_2$ (K-doped Ba122) and
Ba(Fe$_{0.92}$Fe$_{0.08}$)$_2$As$_2$ (Co-doped Ba122) in which no
trace of large horizontal nodal lines is discovered. In this
paper, we present the temperature and angle dependent resistivity
in Ba(Fe$_{0.75}$Ru$_{0.25}$)$_2$As$_2$ single crystals in
different magnetic fields. Our results indicate that the
electronic anisotropy is quite similar among the three systems,
K-doped, Co-doped and Ru-doped Ba122.

\begin{figure}
\includegraphics[scale=0.7]{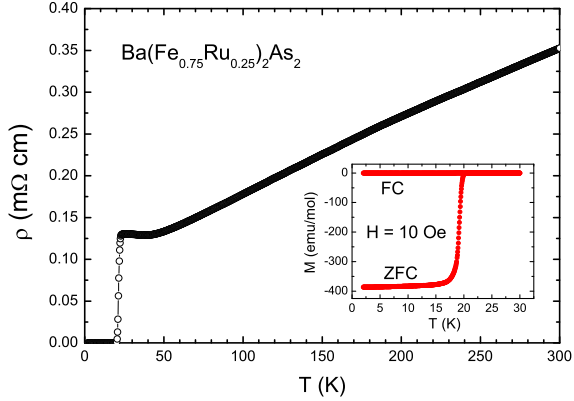}
\caption{(color online) Main panel: The temperature dependence of
resistivity for the Ba(Fe$_{0.75}$Ru$_{0.25}$)$_2$As$_2$ single
crystal in zero field. Inset: The temperature dependent DC
magnetization of the sample. }
\end{figure}

The single crystals of Ba(Fe$_{0.75}$Ru$_{0.25}$)$_2$As$_2$ were
grown with the self-flux method\cite{Kim,Canfield}. The precursor
materials FeAs and RuAs were made from Fe powder (99.9$\%$, Alfa
Aesar), As flakes (99.9$\%$, Alfa Aesar) and Ru filaments
(99.9$\%$, Alfa Aesar). These starting materials were weighed and
mixed in the glover box filled with Ar gas (the oxygen and water
content below 0.1 PPM) and heated to 700 $^\circ$C for 24 hours.
Then the Ba (99.9$\%$, Alfa Aesar), FeAs and RuAs were weighed
according to the ratio Ba:Fe:Ru:As = 1:3:1:4 and mixed well, put
in a quartz tube which is sealed in an iron tube and annealed at
about 1200 $^\circ$C for 20 minutes. In order to protect the iron
tube from being oxidized, an additional quartz tube is shielded
outside the iron tube. A slow cooling down with a rate of 2
$^\circ$C/hour of the iron-tube was taken from 1200 $^\circ$C to
1010 $^\circ$C, and then the power of the furnace was turned off.
The crystals with black and shiny surface and about 3 mm $\times$
3 mm in dimensions are cleaved from the grown-mixture. The XRD
diffraction patterns show only the (00\emph{l}) peaks with a very
narrow full-width at the half-maximum (FWHM). The true composition
of the single crystal was checked with the Energy-dispersive X-ray
spectroscopy (EDX) and found to be close to the nominal one. The
resistivity measurements were taken on a Quantum Design physical
property measurement system (QD, PPMS) with the magnetic fields up
to 14 T. The angle dependent resistivity was measured with a
horizontal rotator (HR) on PPMS. The temperature dependence of
magnetization was measured by a SQUID-VSM (Quantum Design) in the
zero-field-cooling mode (ZFC) and the field-cooling mode (FC). The
temperature dependence of resistivity and magnetization are shown
in Fig.~1. The onset of superconducting transition is found to be
22 K in the $\rho(T)$ curve, with a transition width $\Delta T_c$
$\approx$ 1.6 K (10$\%$ to 90 $\%$ $\rho_n$). The little upturn of
the resistivity just above T$_c$ suggests that our sample resides
at a slight underdoping level. The DC magnetization measured at a
magnetic field of 10 Oe exhibits a sharp transition and shows a
full volume of Meissner screening at 2 K.

\begin{figure}
\includegraphics[scale=0.95]{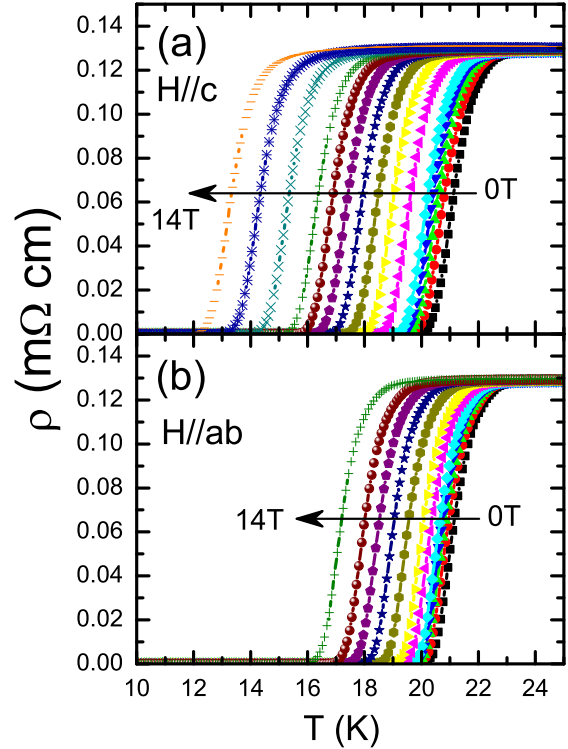}
\caption{(color online)  (a) The  in-plane resistivity of
Ba(Fe$_{0.75}$Ru$_{0.25}$)$_2$As$_2$ single crystal in different
magnetic fields $H$= 0,0.25,0.5,0.75,1,2,3,4,5,6,7,8,10,12,14 T
for $H\parallel c$.  (b) The in-plane resistivity of
Ba(Fe$_{0.75}$Ru$_{0.25}$)$_2$As$_2$ single crystal in different
magnetic fields $H$= 0,0.25,0.5,0.75,1,2,3,4,5,,7,9,11,14 T for
$H\parallel ab$. }
\end{figure}

Fig.~2 shows the temperature dependence of the resistivity
measured in different magnetic fields for both $H\parallel ab$ and
$H\parallel c$. One can see a systematic suppression of the
transition temperature in the magnetic fields. As seen in other
iron-based superconductors, the $T_c$ is suppressed more rapidly
for $H\parallel c$ than for $H\parallel ab$, indicating a higher
upper critical field $H_{c2}$ for $H\parallel ab$. This parallel
shift of the resistive curves under magnetic fields suggest that
the superconducting fluctuation is weak in this sample.

\begin{figure}
\includegraphics[scale=0.75]{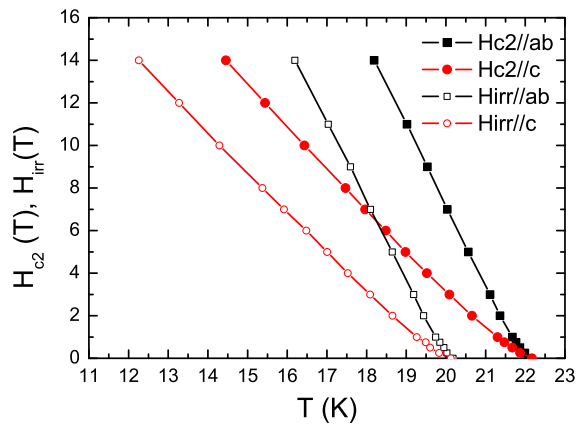}
\caption{(color online)  The upper critical field and the irreversible field of
Ba(Fe$_{0.75}$Ru$_{0.25}$)$_2$As$_2$ single crystal for H $\parallel$ c
and H $\parallel$ ab, respectively. The criterions for determining $H_{C2}$ and $H_{irr}$ are 90\%$\rho_n$ and 1\%$\rho_n$, respectively.}
\end{figure}

Taking the criterions of 90\% $\rho_n$ and 1\% $\rho_n$, we
determined the value of the upper critical field $H_{c2}$ and the
irreversible field $H_{irr}$ respectively, the results are shown
in Fig.~3. The $H_{c2}(T)$ curves show a roughly linear
$T$-dependent behavior, yielding the slopes
$-dH^{ab}_{c2}/dT_c|_{T_c}$ = 3.55 T/K for $H \parallel ab$ and
$-dH^{c}_{c2}/dT_c|_{T_c}$ = 1.82 T/K for $H\parallel c$. These
slopes are generally smaller than those in
Ba$_{0.6}$K$_{0.4}$Fe$_2$As$_2$ and
Ba(Fe$_{0.92}$Co$_{0.08}$)$_2$As$_2$, suggesting a lower upper
critical field in the Ru-doped samples. According to the
Werthamer-Helfand-Hohenberg (WHH)\cite{WHH} formula
$H_{c2}=-0.69(dH_{c2}/dT)|_{T_c}T_c$, the calculated upper
critical fields are $H^{ab}_{c2}(0)$ = 54.3 T and $H^{c}_{c2}(0)$
= 27.8 T. The anisotropy ratio determined here is
$\Gamma$=$H^{ab}_{c2}(0)/H^{c}_{c2}(0)$ = 54.3/27.8 = 1.95. The
values of $H_{c2}$ in the present system is relatively small
compared to others in the family of iron-based superconductors,
but the $\Gamma$ is quite similar among all the Ba122 systems (see
below)\cite{Reid}. As we know, the Pauli limit for singlet pairing
is determined by $H_{c2}(0)/k_BT_c$ = 1.84 T/K\cite{Paulilimit}
when the spin-orbital coupling is weak, therefore the upper
critical fields may be limited by the Pauli limit $H_{c2}^{Pauli}
\approx 36 $ T at low temperatures. It has been pointed out that
the WHH formula, proposed for conventional s-wave superconductor,
may not apply for the iron-based superconductors, which have been
proved to be governed by an unconventional superconducting
mechanism\cite{Schultz}. However, the value of $-dH^{ab}_{c2}/dT$
is proportional to $(m_{ab}\times m_c)^{1/2}$, while
$-dH^{c}_{c2}/dT\propto m_{ab}$, therefore the ratio determined by
$dH^{ab}_{c2}/dT_c|_{T_c}$/$dH^{c}_{c2}/dT_c|_{T_c}$ $\approx$
1.95 is reliably telling the anisotropy of the averaged electron
mass near the transition temperature. In Fig.~3 we also present
the irreversibility lines $H^{ab}_{irr}(T)$ and $H^c_{irr}(T)$,
taking the criterion of 1\%$\rho_n$. One can see that the two
lines, $H_{c2}(T)$ and $H_{irr}(T)$), are parallel to each other.
The area between the lines of $H_{c2}(T)$ and $H_{irr}(T)$ is
small for both cases of $H\parallel c$ and $H\parallel ab$,
indicating a quite weak vortex thermal fluctuation effect. This is
well associated with the small anisotropy factor $\Gamma=1.95$.

\begin{figure}
\includegraphics[scale=0.7]{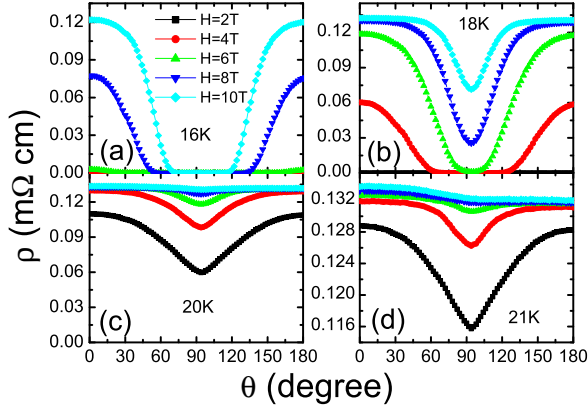}
\caption{(color online) Angular dependence of resistivity at (a)
16 K, (b) 18 K, (c) 20 K and (d) 21 K in $H = $ 2, 4,
 6,  8, 10 T for the Ba(Fe$_0.75$Ru$_0.25$)$_2$As$_2$ single crystal.}
\end{figure}

\begin{figure}
\center
\includegraphics[scale=0.75]{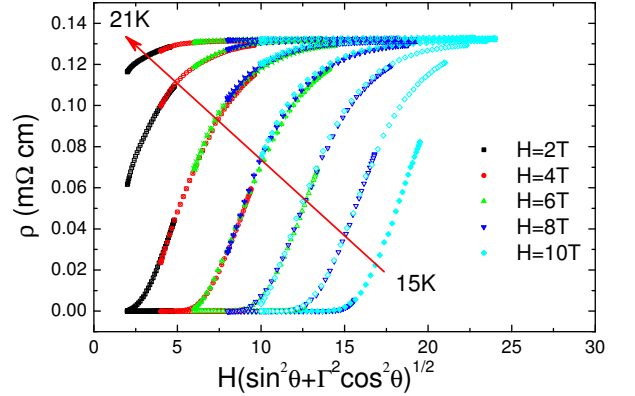}
\caption{(color online) Scaling of the resistivity versus $\tilde{H}=H\sqrt{\sin^2(\theta)+\Gamma^{2}\cos^2(\theta)}$ at 15, 16, 17, 18, 19, 20, 21 K in different magnetic fields. For each temperature, $\Gamma$ is adjusted to achieve the best scaling.}
\end{figure}

In order to further determine the anisotropy $\Gamma$, we measured
the angle-dependent resistivity of
Ba(Fe$_{0.75}$Ru$_{0.25}$)$_2$As$_2$ in different magnetic fields
at several fixed temperatures. The typical results are shown in
Fig.~4. The $\theta$ is the angle that encloses between the
magnetic field and c-axis. During the measurements, the magnetic
field is always applied perpendicular to the direction of the
applied current. Therefore $\theta=0^\circ$ and $\theta=90^\circ$
represent the cases of $H
\parallel c$ and $H \parallel ab$, respectively. According to the
anisotropic Ginzburg-Landau theory, the angle dependent upper
critical field is given by
\begin{equation}
H_{c2}^{GL}(\theta) =
H_{c2}^{c}/\sqrt{\sin^2(\theta)+\Gamma^{2}\cos^2(\theta)}.
\end{equation}
with the anisotropy
\begin{equation}
\Gamma=H_{c2}^{ab}/H_{c2}^{c}=(m_c/m_{ab})^{1/2}=\xi_{ab}/\xi_c.
\end{equation}

It has been proposed by Blatter et al.\cite{Blatter} that, if we
take the x-coordinate as
$\tilde{H}=H\sqrt{\sin^2(\theta)+\Gamma^{2}\cos^2(\theta)}$, and
re-plot the data as $\rho$ vs. $\tilde{H}$, the curves for
different magnetic field at the same temperature will collapse
into one. This scaling law can be used to determine the anisotropy
factor $\Gamma$ at each temperature. Fig.~5 shows the scaled
results for temperatures from 15 K to 22 K. One can see a good
scaling for all temperatures. And $\Gamma$ is optimized for each
temperature to achieve the best scaling. The value of $\Gamma$
determined from the scaling is about 2, which is consistent with
the value derived from the upper critical field, as discussed
above. And $\Gamma$ shows a temperature dependent variation. It
increases monotonically from 1.95 at 15 K to 2.4 at 20 K, and then
drops slightly to 2.25 at 21 K.

In Fig.~6, we show a collection of the anisotropy for the K-doped,
Co-doped, Ru-doped Ba122 and
RbFe$_2$Se$_2$\cite{wangzhaosheng,lichunhong}. It is found that
the absolute value of $\Gamma$ and its temperature dependence in
the Ru-doped sample is quite similar to that of Co-doped and
K-doped Ba122 samples. Regarding this similarity, it is hard to
believe that the warping effect in the Ru-doped sample is much
stronger than that in its counterparts Co-doped and K-doped Ba122
samples. Therefore the possible nodal gap in the Ru-doped samples
may arise from other effect, instead of the strong Fermi surface
warping effect. Recently, the ARPES
measurements\cite{Kaminski,DingH} indicates that the Fermi surface
evolves slowly, and the Fermi velocity $v_F$ (at $k_z=\pi$) seems
also weakly changed up to the optimal doping point (at around x =
0.3) in Ba(Fe$_{1-x}$Ru$_x$)$_2$As$_2$. However, towards more
doping of Ru in the underdoped region, the AF state is strongly
suppressed and superconductvity starts to appear at x = 0.15 and
gets the optimized $T_c$ at about x = 0.30. Clearly it is hard to
understand that the superconductivity is purely induced by the
effect of the Fermi surface nesting. Furthermore, it is also quite
difficult to understand why the anisotropy in the three different
systems are similar to each other, since Co-doped sample has
larger electron Fermi surfaces, while the K-doped one has larger
hole pockets, and the Ru-doped sample is close to the case of
iso-valent doping. To unravel this puzzle, it is highly desired to
do further more elegant band structure calculations, and more
detailed ARPES measurements.

\begin{figure}
\center
\includegraphics[scale=0.75]{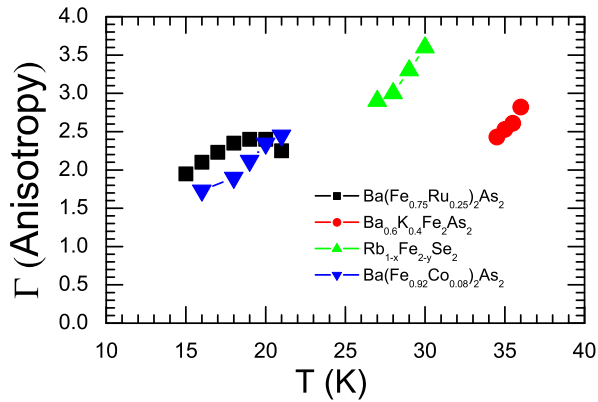}
\caption{(color online) Temperature dependent anisotropy $\Gamma$
of Ba(Fe$_{0.75}$Ru$_{0.25}$)$_2$As$_2$,
Ba$_{0.6}$K$_{0.4}$Fe$_2$As$_2$, Rb$_{1-x}$Fe$_{2-y}$Se$_2$ and
Ba(Fe$_{0.92}$Co$_{0.08}$)$_2$As$_2$ single crystals.}
\end{figure}

In summary, we have measured the temperature and angle dependent
resistivity for Ba(Fe$_{0.75}$Ru$_{0.25}$)$_2$As$_2$ single
crystals in magnetic fields up to 14 T for $H \parallel c$ and $H
\parallel ab$. The calculated upper
critical fields are $H^{ab}_{c2}(0)$ = 54.3 T and $H^{c}_{c2}(0)$
= 27.8 T, being smaller than that in other Ba122 iron-based
superconductors. Interestingly, the anisotropy $\Gamma \approx$
1.95 is similar to the optimally Co-doped or K-doped Ba122. The
anisotropy is also rechecked with an independent way, that is to
scale the resistivity vs. angle curves measured at different
magnetic fields but at a fixed temperature. Both methods yield
similar anisotropy. The similar value of anisotropy found in the
Co-doped, K-doped and the Ru-doped Ba122 systems suggest that the
3D warping effect of the Fermi surface may not be the cause for
the appearance of the nodal gap.

\begin{acknowledgments}
This work is supported by the NSF of China
(11034011/A0402), the Ministry of Science and Technology of China
(973 projects: 2011CBA00102 and 2012CB821403) and PAPD.
\end{acknowledgments}


\begin{thebibliography}{10}


\bibitem{Mazin} I. I. Mazin, D. J. Singh, M. D. Johannes, and M. H. Du, Phys. Rev. Lett. {\bf101}, 057003 (2008).
\bibitem{Kuroki} K. Kuroki, S. Onari, R. Arita, H. Usui, Y. Tanaka, H. Kontani, and H. Aoki, Phys. Rev. Lett. {\bf101}, 087004 (2008).
\bibitem{Hanaguri} T. Hanaguri, S. Niitaka, K. Kuroki, and H.
Takagi, Science \textbf{328}, 474(2010).
\bibitem{Christianson} A. D. Christianson, E. A. Goremychkin, R. Osborn, S. Rosenkranz, M. D. Lumsden, C. D. Malliakas, I. S. Todorov, H. Claus, D. Y. Chung, M. G. Kanatzidis, R. I. Bewley, and T. Guidi, Nature \textbf{456}, 930-932 (2008).
\bibitem{Keimer} D. S. Inosov, J. T. Park, P. Bourges, D. L. Sun, Y. Sidis, A. Schneidewind, K. Hradil, D. Haug, C. T. Lin, B. Keimer, and  V. Hinkov, Nature Physics \textbf{6}, 178 - 181 (2010).
\bibitem{ImaiNMR} F. L. Ning, K. Ahilan, T. Imai, A. S. Sefat, M. A. McGuire, B. C. Sales, D. Mandrus, P. Cheng, B. Shen, and H. -H. Wen, Phys. Rev.
Lett. \textbf{104}, 037001 (2010).
\bibitem{IshidaNMR} Y. Nakai, T. Iye, S. Kitagawa, K. Ishida, H. Ikeda, S. Kasahara, H. Shishido, T. Shibauchi, Y. Matsuda, and T. Terashima
, Phys. Rev. Lett. {\bf105}, 107003 (2010).
\bibitem{GraserPRB2009} V. Mishra, G. Boyd, S. Graser, T. Maier, P. J. Hirschfeld, and D. J. Scalapino, Phys. Rev. B \textbf{79}, 094512 (2009).
\bibitem{KurokiPRB}Kazuhiko Kuroki, Hidetomo Usui, Seiichiro Onari, Ryotaro Arita,
Hideo Aoki, Phys. Rev. B {\bf79}, 224511 (2009).
\bibitem{HirschfeldPRB2010} S. Graser, A. F. Kemper, T. A. Maier, H.-P. Cheng, P. J. Hirschfeld, and D. J. Scalapino, Phys. Rev. B \textbf{81}, 214503 (2010).
\bibitem{HirschfeldReview} P. J. Hirschfeld, M.M. Korshunov, and I.I. Mazin, Reports on Progress in Physics 74, 124508 (2011).
\bibitem{Hicks} Clifford W. Hicks, Thomas M. Lippman, Martin E. Huber, James G. Analytis, Jiun-Haw Chu, Ann S. Erickson, Ian R. Fisher, and Kathryn A. Moler
, Phys. Rev. Lett. {\bf103}, 127003 (2009).
\bibitem{Carritong} J. D. Fletcher, A. Serafin, L. Malone, J. G. Analytis, J.-H. Chu, A. S. Erickson, I. R. Fisher, and A. Carrington, Phys. Rev. Lett. {\bf102}, 147001 (2009).
\bibitem{LiSYKFe2As2} J. K. Dong, S. Y. Zhou, T. Y. Guan, H. Zhang, Y. F. Dai, X. Qiu, X. F. Wang, Y. He, X. H. Chen, and S. Y. Li, Phys. Rev. Lett. {\bf104}, 087005 (2010).
\bibitem{ZengB} B. Zeng, G. Mu, H. Q. Luo, T. Xiang, I. I. Mazin, H. Yang, L. Shan, C. Ren, P. C. Dai, and H. H. Wen, Nat. Commun. {\bf1}, 112 (2010).
\bibitem{MatsudaPDoping} K. Hashimoto, M. Yamashita, S. Kasahara, Y. Senshu, N. Nakata, S. Tonegawa, K. Ikada, A. Serafin, A. Carrington, T. Terashima, H. Ikeda, T. Shibauchi, and Y. Matsuda, Phys. Rev.
B \textbf{81}, 220501(R) (2010) .
\bibitem{FengDLPDoping} Y. Zhang, Z. R. Ye, Q. Q. Ge, F. Chen, Juan Jiang, M. Xu, B. P. Xie, and D. L. Feng, accepted by Nature Physics, (2012).
\bibitem{Sharma} S. Sharma, A. Bharathi, S. Chandra, R. Reddy, S. Paulraj, A. Satya, V. Sastry, A. Gupta, and C. Sundar, Phys. Rev. B \textbf{81}, 174512 (2010).
\bibitem{Thaler} A. Thaler, N. Ni, A. Kracher, J. Q. Yan, S. L. Bud¡¯ko, and P. C. Canfield, Phys. Rev. B \textbf{82}, 014534 (2010).
\bibitem{Goldman} M. G. Kim, D. K. Pratt, G. E. Rustan, W. Tian, J. L. Zarestky, A. Thaler, S. L. Bud¡¯ko, P. C. Canfield, R. J. McQueeney, A. Kreyssig, and A. I. Goldman, Phys. Rev. B \textbf{83}, 054514 (2011).
\bibitem{Poissonnet} F. Rullier-Albenque, D. Colson, A. Forget, P. Thu\'ery, and S. Poissonnet, Phys. Rev. B \textbf{81}, 224503 (2010).
\bibitem{LiSY} X. Qiu, S. Y. Zhou, H. Zhang, B. Y. Pan, X. C. Hong, Y. F. Dai, Man Jin Eom, Jun Sung Kim, Z. R. Ye, Y. Zhang, D. L. Feng, and S. Y. Li, Phys. Rev. X \textbf{2}, 011010 (2012).
\bibitem{Kaminski} R. S. Dhaka, Chang Liu, R. M. Fernandes, Rui Jiang, C. P. Strehlow, Takeshi Kondo, A. Thaler, Jorg Schmalian, S. L. Bud'ko, P. C. Canfield, Adam Kaminski, Phys. Rev. Lett. \textbf{107}, 267002 (2011).
\bibitem{Prozorov} R. T. Gordon, N. Ni, C. Martin, M. A. Tanatar, M.D. Vannette, H. Kim, G. D. Samolyuk, J. Schmalian, S. Nandi, A. Kreyssig, A. I. Goldman, J. Q. Yan, S. L. Bud¡¯ko, P. C. Canfield, and R. Prozorov, Phys. Rev. Lett. {\bf102}, 127004 (2009).
\bibitem{Reid} J.-Ph. Reid, M. A. Tanatar, X. G. Luo, H. Shakeripour, N. Doiron-Leyraud, N. Ni, S. L. Bud¡¯ko, P. C. Canfield, R. Prozorov, and Louis Taillefer, Phys. Rev. B \textbf{82}, 064501 (2010).
\bibitem{Kim} M. J. Eom, S. W. Na, C. Hoch, R. K. Kremer, and J. S. Kim, arXiv:1109.1083 (2011).
\bibitem{Canfield} N. Ni, M. E. Tillman, J.-Q. Yan, A. Kracher, S. T. Hannahs, S. L. Bud¡¯ko, and P. C. Canfield, Phys. Rev. B \textbf{78}, 214515 (2008).
\bibitem{WHH} N. R. Werthamer, E. Helfand, and P. C. Hohenberg, Phys. Rev. \textbf{147}, 295 (1966).
\bibitem{Paulilimit} A. M. Clogston, Phys. Rev. Lett. \textbf{9}, 266 (1962).
\bibitem{Schultz} G. Fuchs, S.-L. Drechsler, N. Kozlova, G. Behr, A. Koehler, J. Werner, K. Nenkov, C. Hess, R. Klingeler, J.E. Hamann-Borrero, A. Kondrat, M. Grobosch, M. Knupfer, J. Freudenberger, B. Buechner, and L.
Schultz, Phys. Rev. Lett. \textbf{101}, 237003 (2008).
\bibitem{Blatter} G. Blatter, V. B. Geshkenbein, and A. I. Larkin,  Phys. Rev. Lett. \textbf{68}, 875 (1992).
\bibitem{wangzhaosheng} Zhao-Sheng Wang, Hui-Qian Luo, Cong Ren, and Hai-Hu Wen, Phys. Rev. B \textbf{78}, 140501 (2008).
\bibitem{lichunhong} Chun-Hong Li, Bing Shen, Fei Han, Xiyu Zhu, and Hai-Hu Wen, Phys. Rev. B \textbf{83}, 184521 (2011).
\bibitem{DingH}  N. Xu, T. Qian, P. Richard, Y.-B. Shi, X.-P. Wang, P. Zhang, Y.-B. Huang, Y.-M. Xu, H. Miao, G. Xu, G.-F. Xuan, W.-H. Jiao, Z.-A. Xu, G.-H. Cao, and H. Ding,  arXiv:1203.4699.


\end{thebibliography}
\end{document}